\documentclass[useAMS,usenatbib]{mn2e}

\usepackage{graphicx}
\usepackage{wasysym}
\usepackage{amssymb}
\usepackage{stmaryrd}

\title[WASP-5b: a transiting very-hot Jupiter]
{WASP-5b: a dense, very-hot Jupiter
transiting a 12th-mag Southern-hemisphere star}

\author[D.~R.~Anderson et al.]{D.~R.~Anderson$^{1}$\thanks{E-mail:
dra@astro.keele.ac.uk}, M.~Gillon$^{2,3}$, C.~Hellier$^{1}$, P.~F.~L.~Maxted$^{1}$, F.~Pepe$^{2}$, \newauthor D.~Queloz$^{2}$, D.~M.~Wilson$^{1}$, A.~Collier~Cameron$^{4}$, B.~Smalley$^{1}$, T.~A.~Lister$^{1,5}$, \newauthor S.~J.~Bentley$^{1}$, A.~Blecha$^{2}$,
D.~J.~Christian$^{6}$, B.~Enoch$^{7}$, L.~Hebb$^{4}$, K.~Horne$^{4}$, \newauthor J.~Irwin$^{8}$, Y.~C.~Joshi$^{6}$, S.~R.~Kane$^{9}$, M.~Marmier$^{2}$, M.~Mayor$^{2}$, N.~R.~Parley$^{7}$, \newauthor D.~L.~Pollacco$^{6}$, F.~Pont$^{2}$, R.~Ryans$^{6}$, D.~S\'egransan$^{2}$, I.~Skillen$^{10}$, R.~A.~Street$^{5,6,11}$, \newauthor S.~Udry$^{2}$, R.~G.~West$^{12}$, P.~J.~Wheatley$^{13}$\\
$^{1}$Astrophysics Group, Keele University, Keele, ST5 5BG, UK\\
$^{2}$Observatoire de Gen\`eve, 51 ch. des Maillettes, 1290 Sauverny, Switzerland\\
$^{3}$Institut d'Astrophysique et de G\'eophysique,  Universit\'e de
Li\`ege,  4000 Li\`ege, Belgium \\
$^{4}$School of Physics and Astronomy, University of St. Andrews, North Haugh, Fife, KY16 9SS, UK\\
$^{5}$Las Cumbres Observatory, 6740 Cortona Dr. Suite 102, Santa Barbara, CA 93117, USA\\
$^{6}$Astrophysics Research Centre, School of Mathematics \& Physics, Queen's University, University Road, Belfast, BT7 1NN, UK\\
$^{7}$Department of Physics and Astronomy, The Open University, Milton Keynes, MK7 6AA, UK\\
$^{8}$Department of Astronomy, Harvard University, 60 Garden Street, MS 10, Cambridge, Massachusetts 02138, USA\\
$^{9}$Michelson Science Center, Caltech, MS 100-22, 770 South Wilson Avenue,
Pasadena, CA 91125, U.S.A\\
$^{10}$Isaac Newton Group of Telescopes, Apartado de Correos 321, E-38700 Santa Cruz de la Palma, Tenerife, Spain\\
$^{11}$Department of Physics, Broida Hall, University of California, 
Santa Barbara, CA 93106-9530, USA\\
$^{12}$Department of Physics and Astronomy, University of Leicester, Leicester, LE1 7RH, UK\\
$^{13}$Department of Physics, University of Warwick, Coventry, CV4 7AL, UK}

\begin{document}

\date{Accepted YYYY Month DD. Received YYYY Month DD; in original form YYYY Month DD}

\pagerange{\pageref{firstpage}--\pageref{lastpage}} \pubyear{2007}

\maketitle

\label{firstpage}

\begin{abstract}
We report the discovery of WASP-5b, a Jupiter-mass
planet orbiting a 12th-mag G-type star in the Southern
hemisphere.  The 1.6-d orbital period places WASP-5b
in the class of very-hot Jupiters and leads to a
predicted equilibrium temperature of 1750 K. WASP-5b
is the densest of any known Jovian-mass planet, being a
factor seven denser than TrES-4, which is subject to 
similar stellar insolation, and a factor three denser 
than WASP-4b, which has a similar orbital period. 
We present transit photometry and
radial-velocity measurements of 
WASP-5 (= USNO-B1\,0487-0799749), from which we 
derive the mass, radius and density
of the planet: M$_{\rm P}$ = 1.58 $^{+0.13}_{-0.08}$
M$_{\rm J}$, R$_{\rm P}$ = 1.090 $^{+0.094}_{-0.058}$ 
R$_{\rm J}$ and $\rho_{\rm P}$ = 1.22 $^{+ 0.19}_{- 0.24}$
 $\rho_{\rm J}$.
The orbital period is P = 1.6284296 
$^{+0.0000048}_{-0.0000037}$ d and the
mid-transit epoch is T$_{\rm C}$ (HJD) = 2454375.62466
$^{+0.00026}_{-0.00025}$.
\end{abstract}

\section{INTRODUCTION}
Several wide-angle, ground-based surveys for transiting extrasolar
planets have now been successful, including HAT
\citep{2002PASP..114..974B}, TrES \citep{2006ApJ...651L..61O}, WASP
\citep{2006PASP..118.1407P} and XO \citep{2005PASP..117..783M}.  In
the Southern hemisphere relatively faint ($V$ = 16--17) transiting
systems have been found by OGLE
\citep{1992AcA....42..253U,2007arXiv0710.5278P} and by \citet{2007arXiv0711.1746W}. The WASP-South survey is
the first to find much brighter transiting systems in the South
\citep{WASP-4b}. We report here the second planet found by the
WASP-South and CORALIE collaboration, a dense, very-hot Jupiter transiting a
$V$ = 12.3 star with an orbital period of 1.6 d.

Transit searches are most sensitive to large, Jupiter-sized planets 
in very short orbits. WASP-5b, like the recently announced WASP-3b 
and WASP-4b, has an ultra-short 
period below 2 d \citep{WASP-3b,WASP-4b}.  We discuss the characteristics of these 
ultra-close, highly irradiated planets.

\section{OBSERVATIONS}
The WASP-South observatory is sited at the Sutherland station
of the South African Astronomical Observatory. It consists 
of an array of 8 cameras, each with a Canon 200-mm 
f/1.8 lens backed by an
$\itl e2v$ 2k$\times$2k CCD, covering 7.8$^{\circ}\times$7.8$^{\circ}$.
WASP-South started operating in May 2006 with a strategy of tiling
up to eight pointings with a cadence of 5--10 mins, taking two
30-s exposures at each pointing. The WASP hardware, observing 
strategy, data-reduction pipeline and archive are described in 
detail by \citet{2006PASP..118.1407P}.   The data
were detrended and transit-searched as described in
\citet{2006MNRAS.373..799C}, and candidates were selected as described in
\citet{2007MNRAS.380.1230C}.

\begin{figure}
\includegraphics[width=84mm, angle=270]{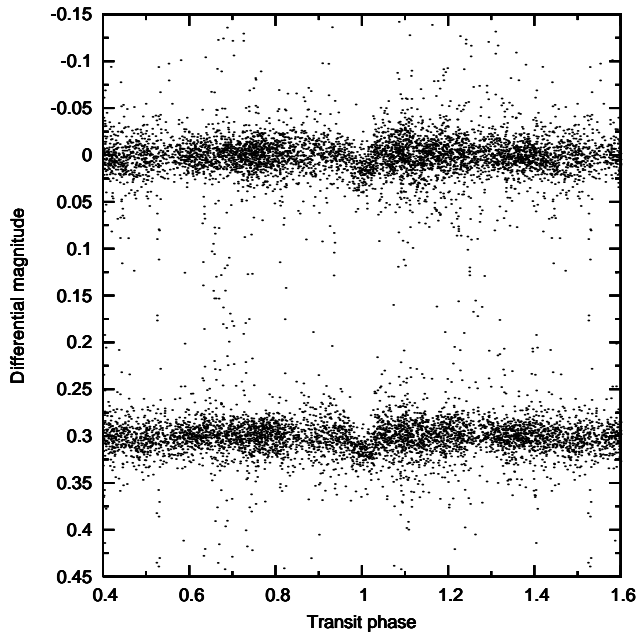} 
\caption{Discovery light curves for WASP-5 obtained with WASP-South
camera 7 (top) and WASP-South camera 8 (bottom, offset by 0.3 mag for 
clarity).}
\label{SW-phot}
\end{figure}

WASP-5 (= 1SWASP\,J235723.74--411637.5 = USNO-B1\,0487-0799749 =
2MASS\,23572375--4116377 = NOMAD1\,0487-0881175) is listed 
in the NOMAD catalogue 
as a $V$ = 12.3 star, and was observed by WASP-South from 2006
May 13 to 2006 November 14.  Transits of WASP-5b were detected by two
WASP-South cameras, where they overlap at the field edges. A total of
4642 data points were obtained with camera 7 and 4108 data points with
camera 8 (Fig.~\ref{SW-phot}).

\begin{figure}
\includegraphics[width=90mm, angle=270]{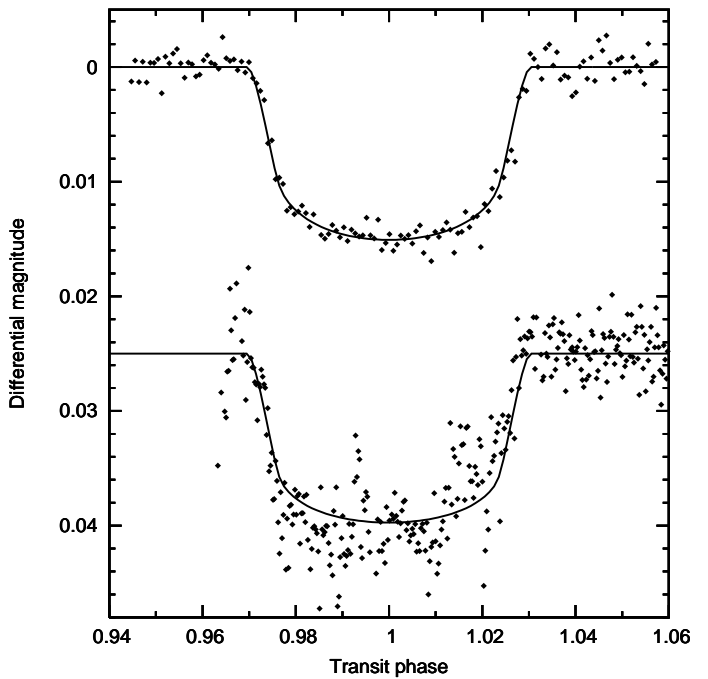} 
\caption{The $R$-band light curve obtained with the Euler telescope (top) and
the SDSS $i$'-band  light curve obtained with  Faulkes Telescope South (bottom, offset by 0.025 mag for clarity).
In each case the solid line is the optimal MCMC solution (see \S \ref{MCMC}).}
\label{followup-phot}
\end{figure}

During follow-up, a transit was observed in the $\itl R$-band using
EulerCam on the 1.2-m telescope at La Silla on 2007 October 10
(Fig.~\ref{followup-phot}). The
telescope was heavily defocussed to allow long exposure times
(2 min) resulting in an RMS scatter of 1-mmag.  A further
transit was observed in the SDSS $\itl i'$ band on 2007 October 13
using HawkCam2 on the 2.0-m Faulkes Telescope South at Siding Spring
Observatory (Fig.~\ref{followup-phot}).

Radial-velocity data were obtained with the CORALIE spectrograph on
the Euler 1.2-m telescope.  CORALIE had recently been upgraded, as
described in \citet{WASP-4b}. Eleven radial velocity measurements were
taken over the course of one month (Table \ref{rv-data};
Fig.~\ref{RV-multi}), establishing WASP-5b as a planetary-mass
companion.

We used a line-bisector analysis to look for asymmetries 
in the spectral line profiles,
as could be caused by contamination from an unresolved eclipsing binary 
\citep{2001A&A...379..279Q, 2005ApJ...621.1061M}. Such a binary would produce bisector spans that vary in phase with the photometric period with an amplitude comparable to the radial-velocity amplitude. This is not seen in our data (Fig.~\ref{RV-multi}, bottom panel), supporting the conclusion
that the radial-velocity variations are due to a planet.

\begin{table}
 \caption{Radial velocity measurements of WASP-5}
 \label{rv-data}
 \begin{tabular*}{0.5\textwidth}{@{\extracolsep{\fill}}cccc}
 \hline
 BJD--2,400,000 & RV & $\sigma$$_{RV}$ & BS$^{a}$\\
 (days) & (km s$^{-1}$) & (km s$^{-1}$) & (km s$^{-1}$)\\
 \hline
 54359.614570 & 19.785 & 0.016 & --0.023\\
 54362.654785 & 19.942 & 0.015 & \,\,\,0.033\\
 54364.676219 & 19.753 & 0.021 & \,\,\,0.026\\
 54365.682733 & 20.203 & 0.013 & \,\,\,0.015\\
 54372.781658 & 19.750 & 0.013 & --0.047\\
 54374.817984 & 20.016 & 0.040 & \,\,\,0.091\\
 54376.714797 & 20.248 & 0.014 & --0.024\\
 54377.762440 & 19.747 & 0.013 & --0.002\\
 54379.627523 & 19.907 & 0.014 & --0.060\\
 54380.682557 & 19.830 & 0.010 & --0.024\\
 54387.644986 & 19.818 & 0.015 & \,\,\,0.027\\
 \hline
 \end{tabular*}
 \\$^{a}$ Bisector spans; $\sigma$$_{BS}$ $\approx$ 2\,$\sigma$$_{RV}$
\end{table}
\section{STELLAR PARAMETERS}
The CORALIE spectra, when co-added, give a
signal-to-noise ratio of $\sim$\,40, which is
suitable for a preliminary photospheric analysis of WASP-5. The
analysis was performed using the {\sc uclsyn} spectral synthesis
package and {\sc atlas9} models without convective overshooting
\citep*{1997A&A...318..841C}. The H$\alpha$, Na {\sc i} D and Mg {\sc
i} b lines were used as diagnostics of both $T_{\rm eff}$ and $\log
g$. The metallicity was estimated using the photospheric lines in the
6000--6200 \AA\ region. The parameters obtained from this analysis 
are listed in Table~\ref{stellar-params}.

We also used the GALEX NUV flux and magnitudes from the NOMAD, DENIS 
and 2MASS catalogues to estimate the effective temperature using 
the Infrared Flux Method (IRFM)
\citep{1977MNRAS.180..177B}. This gives $T_{\rm eff} = 5610 \pm
250$~K, which is in close agreement with the
spectroscopic analysis. These results imply a spectral type in the
range G2V--G6V. The Li {\sc i} 6708 \AA\ line is not present in the
coadded spectrum, giving an upper limit on the lithium abundance of
$\log A(\rm Li)$ $\lesssim$ 1.2. This implies an age of $\gtrsim$ 
2 Gyr for a star of this effective temperature
\citep{2005A&A...442..615S}. The value of $v \sin i$ = 3.4 km s$^{-1}$ obtained from the CORALIE spectra indicates an age of 1.7--4.4 Gyr \citep{1997MmSAI..68..881B}. Comparison of the temperature and $\log
g_{\rm *}$ with the stellar evolution models of \citet{2000A&AS..141..371G}
gives maximum-likelihood values of $M_{\rm *}$ = 0.99 $\pm$ 0.08
M$_{\rm \astrosun}$ and $R_{\rm *}$ = 0.97 $\pm$ 0.06 R$_{\rm
\astrosun}$. The distance of WASP-5 ($d$ = 300 $\pm$ 50 pc) was
calculated using the distance modulus, the
NOMAD apparent visual magnitude ($V$ = 12.3) and the absolute
visual magnitude of a G4V star \citep[$V$ = 4.9,][]{1992oasp.book.....G}.
We assumed $E(B-V)$ = 0, which is reasonable at such a distance 
and in a direction out of the galactic plane.
\begin{table}
 \caption{Stellar parameters for WASP-5}
 \label{stellar-params}
 \begin{tabular*}{0.5\textwidth}{@{\extracolsep{\fill}}lc}
 \hline
 Parameter & Value\\
 \hline
 R.A. (J2000) & 23$^{h}57^{m}23.74^{s}$\\
 \medskip
 Dec. (J2000) & --41$^{\circ}$16\arcmin37\farcs5\\
 $T_{\rm eff}$ (K) & 5700 $\pm$ 150\\
 $\log g_{*}$ (cgs)     & 4.3 $\pm$ 0.2\\
 $[$M/H$]$         & 0.0 $\pm$ 0.2\\
 $v \sin i$ (km s$^{-1}$) & 3.4 $\pm$ 0.7\\
 \medskip
 $\log A(\rm Li)$  & $<$ 1.0 \\
 $M_{\rm *}$ (M$_{\rm \astrosun}$) & 0.99 $\pm$ 0.08\\
 $R_{\rm *}$ (R$_{\rm \astrosun}$) & 0.97 $\pm$ 0.06\\
 Spec. Type & G4V $\pm$ 2 subtypes\\
 Age (Gyr) & 1.7--4.4\\
 Distance (pc) & 300 $\pm$ 50\\
 \hline
 \end{tabular*}\\
\end{table}
\section{SPECTROSCOPIC ORBITAL SOLUTION \& PLANETARY PARAMETERS}
\label{MCMC}
The CORALIE radial-velocity measurements were combined
with the WASP-South and EulerCAM photometry in a
simultaneous Markov-chain Monte-Carlo (MCMC) analysis to find
the system parameters. This process is described in detail in \citet{2007MNRAS.380.1230C} and 
\citet{WASP-3b}. The FTS data were not used owing to their
higher red noise. The best-fitting solution to the radial-velocity 
data gave an eccentricity of $e$~=~0.03~$\pm$~0.03. 
Our adopted MCMC solution fixes
the eccentricity at $e$~=~0, which is expected for
such a short period planet.

\begin{figure}
\includegraphics[width=60mm, angle=270]{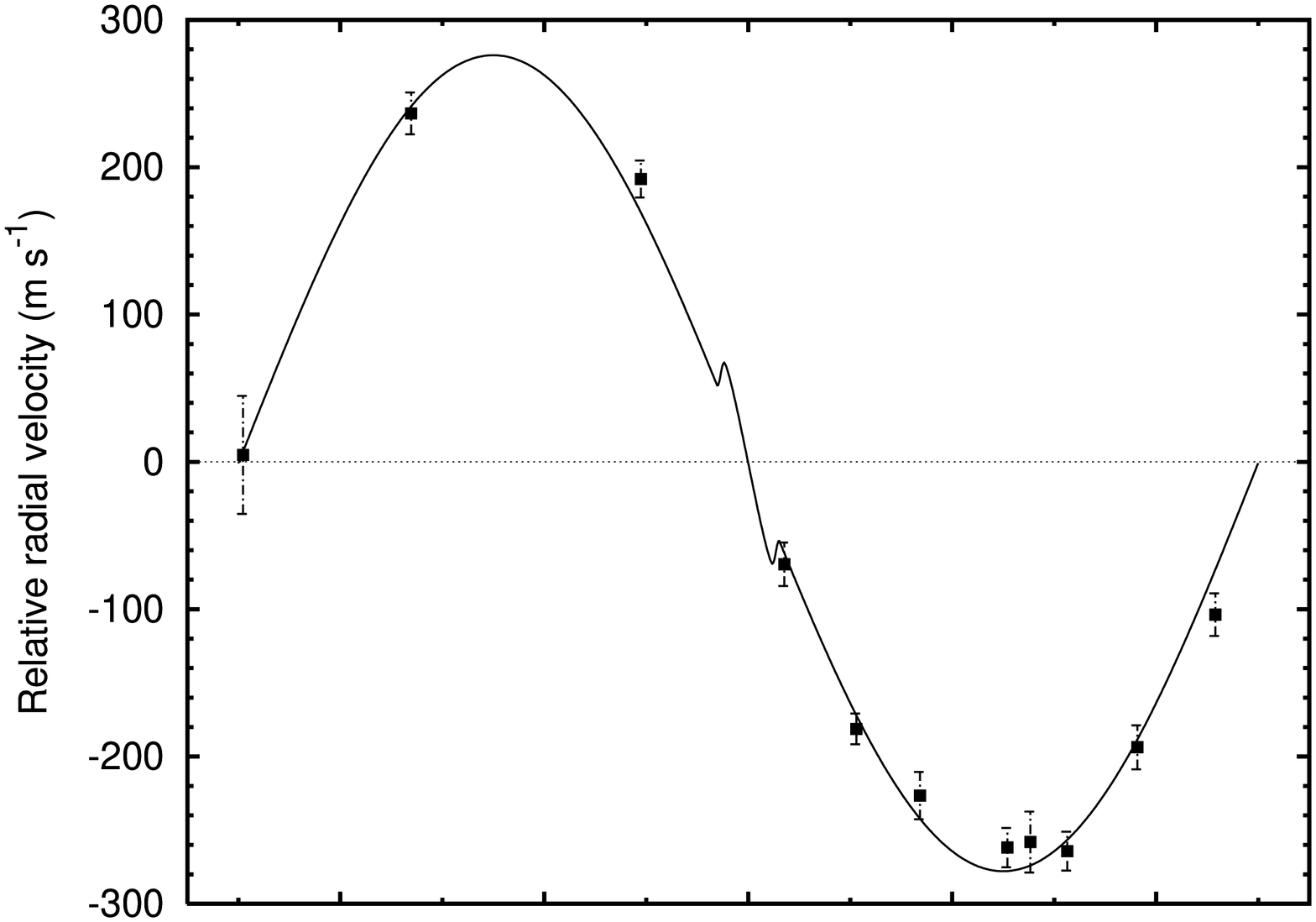} 
\includegraphics[width=59.1mm, angle=270]{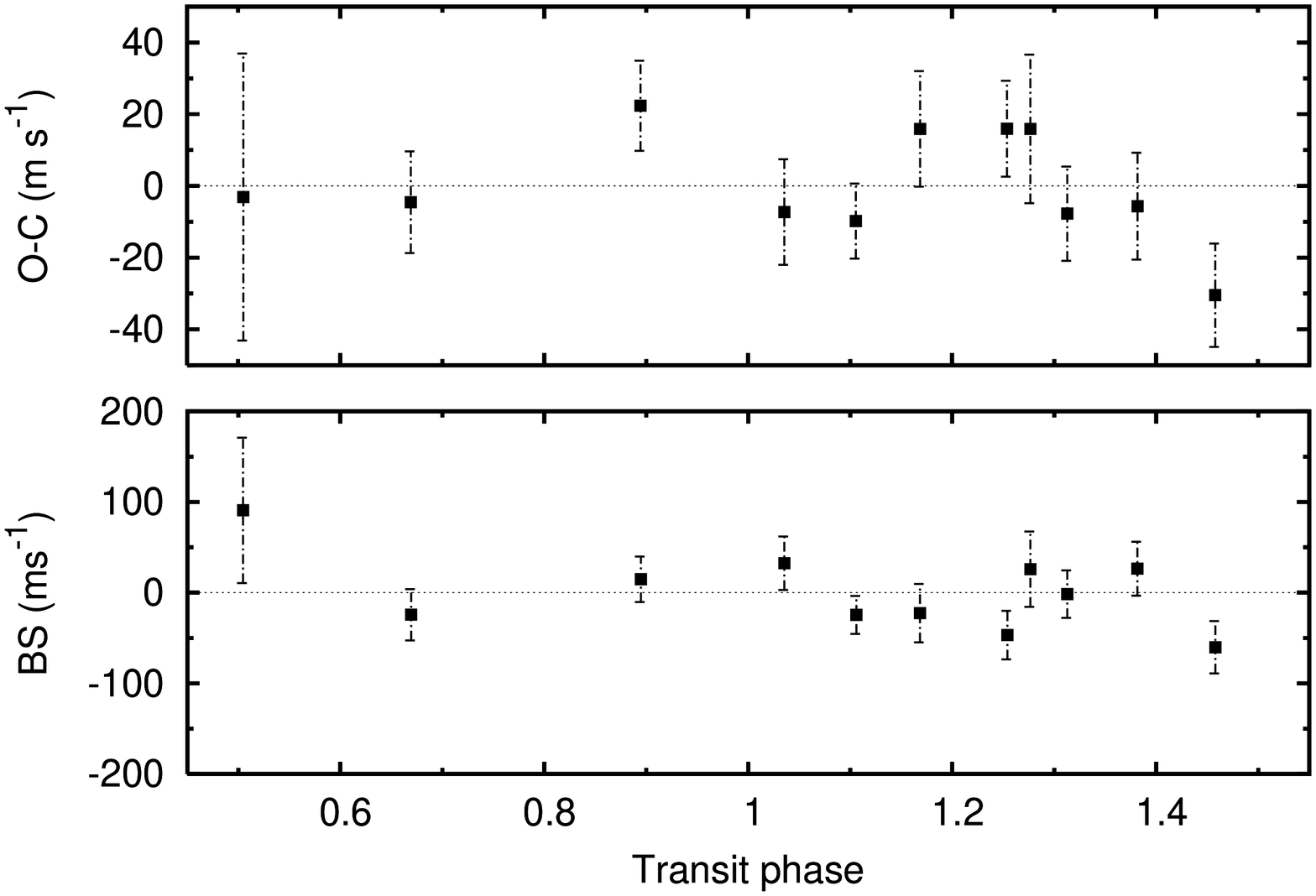} 
\caption{Top: The radial-velocity measurements of WASP-5 obtained
with CORALIE. The centre-of-mass velocity has been subtracted. The
solid line is the MCMC solution (\S \ref{MCMC}); it includes the
Rossiter--McLaughlin effect resulting from a $v \sin i$ of 
3.4 km s$^{-1}$, as obtained from the CORALIE spectra.
Middle: The radial-velocity residuals about the MCMC fit. 
Bottom: Bisector spans for the 11 CORALIE spectra. }
\label{RV-multi}
\end{figure}

In optimising the MCMC solution we used constraints (Gaussian priors)
on the stellar parameters, setting M$_{*}$ = 0.99 $\pm$ 0.08 M$_{\astrosun}$
and $\log g_{*}$ = 4.30 $\pm$
0.20. 
In order to balance the weights of the photometry
and radial velocities in the MCMC analysis, we added
a systematic error of  7 m s$^{-1}$ to the radial
velocities (as might arise, for example, from stellar
activity) to reconcile $\chi^{2}$ with the number
of degrees of freedom; the error bars in 
Fig.~\ref{RV-multi} do not incorporate 
this error.
The optimal MCMC parameters are shown in Table~\ref{sys-params}.

\begin{table}
 \caption{System parameters for WASP-5}
 \label{sys-params}
 \begin{tabular}{lcc}
 \hline
 Parameter & \multicolumn{2}{c}{Value}\\
 \hline
 $P$ (days) & 1.6284296 & $^{+ 0.0000048}_{- 0.0000037}$\\
 $T_{\rm C}$ (HJD) & 2454375.62466 & $^{+ 0.00026}_{- 0.00025}$\\
 $T_{\rm 14}$ (days)$^{a}$ & 0.0979 & $^{+ 0.0011}_{- 0.0009}$\\
 $R_{\rm P}^{2}$/R$_{*}^{2}$ & 0.01192 & $^{+ 0.00038}_{- 0.00027}$\\
 $b$ $\equiv$ $a \cos i/R_{\rm *}$ (R$_{\rm \astrosun}$) & $<$ 0.46 & \\
 $i$ (degs) & $>$ 85.0 & \\
\medskip 
 $e$ & 0 (adopted)\\
 $K_{\rm 1}$ (km s$^{-1}$) & 0.2778 & $^{+ 0.0085}_{- 0.0078}$\\
 $\gamma$ (km s$^{-1}$) & 20.0105 & $^{+ 0.0031}_{- 0.0034}$\\
 $M_{\rm *}$ (M$_{\rm \astrosun}$) & 0.972 & $^{+ 0.099}_{- 0.079}$\\
 $R_{\rm *}$ (R$_{\rm \astrosun}$) & 1.026 & $^{+ 0.073}_{- 0.044}$\\
 $T_{\rm eff}$ (K) & 5883\\
\medskip 
 $\log g_{*}$ (cgs) & 4.403 & $^{+ 0.039}_{- 0.048}$\\
 $M_{\rm P}$ (M$_{\rm J}$) & 1.58 & $^{+ 0.13}_{- 0.08}$\\
 $R_{\rm P}$ (R$_{\rm J}$) & 1.090 & $^{+ 0.094}_{- 0.058}$\\
 $\rho_{\rm P}$ ($\rho_{\rm J}$) & 1.22 & $^{+ 0.19}_{- 0.24}$\\
 $a$ (AU) & 0.02683 & $^{+ 0.00088}_{- 0.00075}$\\
 $\log g_{\rm P}$ (cgs) & 3.484 & $^{+ 0.043}_{- 0.062}$\\
 $T_{\rm P}$ (K) & 1753 & $^{+ 51}_{- 36}$\\
 \hline
\multicolumn{3}{l}{$^{a}$ $T_{\rm 14}$: transit duration, the time between 1$^{st}$ and 4$^{th}$ contact}
 \end{tabular}
\end{table}

\section{Discussion}
\begin{figure}
\includegraphics[width=60mm, angle=270]{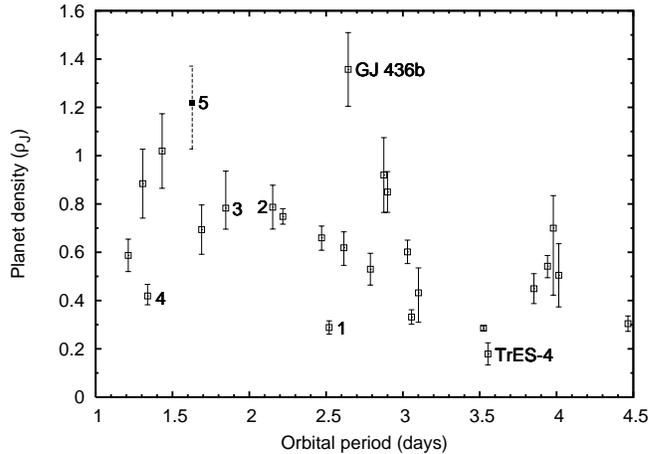} 
\caption{Planet density versus orbital period for planets with 
periods less than 5 d. The dense, hot Neptune, GJ\,436b, and under-dense TrES-4 are labelled; the five WASP planets are labelled numerically.
The error bars on the densities were calculated from the uncertainties on the masses and radii, which were assumed to be uncorrelated. Data taken from http://www.inscience.ch/transits/}
\label{dens-per}
\end{figure}

The discovery of WASP-3b \citep{WASP-3b}, WASP-4b \citep{WASP-4b} and WASP-5b,
all transiting, Jupiter-mass planets with orbital periods of
less than 2 d, increases the number of such systems known from
four to seven\footnote{http://www.inscience.ch/transits/}.
It is notable that they show a large
disparity in density (Fig.~\ref{dens-per}). WASP-5b is the densest
of these systems, whereas WASP-4b is the least dense, with
a density a third that of WASP-5b. This illustrates that planets
with a similar orbital period and radiation environment
must still have different compositions or past histories.

For example, WASP-5b will receive the same insolation and
thus has a similar estimated equilibrium
temperature (1753 K) to TrES-4 (1720 K), which has a
longer orbit about a more luminous star
\citep{2007ApJ...667L.195M}. Despite
this, WASP-5b is denser by a factor of 7
(1.22 $\rho_{J}$ compared with 0.167 $\rho_{J}$ for TrES-4).
\citet{2007ApJ...661..502B} postulate that the bloating of 
the under-dense planets could be due to an enhanced 
atmospheric opacity, which could be absent in WASP-5b.

Although it is denser than other planets with a similar
orbital period, WASP-5b is still within the expected
theoretical range \citep*{2007ApJ...659.1661F}; depending on 
the age of the system, a core mass of 
$\sim$\,50--100 M$_{\varoplus}$ is expected, with a younger 
planet requiring a more massive core.
However, if the bloating of some planets were due to an
extra energy source that is common to all hot Jupiters,
then the core of a dense planet such as WASP-5b would have
to be more massive to compensate 
\citep{2006ApJ...642..495F,2006A&A...453L..21G}.
If the core mass is toward the higher end of the
estimate and the metallicity is no more than solar
then there could be a discrepancy with the
relationship between core mass and stellar
metallicity postulated by \citealt{2007ApJ...661..502B}.
High resolution spectra with better S/N are required 
to reduce the uncertainties in the metallicity and age.

WASP-5b ($\log g_{\rm *}$ = 3.48) has the highest surface gravity
of planets with orbital periods less than 5 d.  Thus it could be
a good system for studying evaporation as a function
of surface gravity. In comparison HD\,209458b is heavily bloated
($\log g_{\rm *}$ = 2.97) and appears to be undergoing evaporation 
\citep{2003Natur.422..143V,2007ApJ...671L..61B,2008arXiv0802.0587V}.
Given its high equilibrium temperature, WASP-5b is also
expected to have a prominent secondary transit, which should be
detectable with $\itl Spitzer$ (e.g. \citealt{2007Natur.447..691H}).  
Further, with the moderately bright host star having a moderate 
rotation rate of $v \sin i = 3.4$ km~s$^{-1}$, the 
Rossiter--McLaughlin effect (40 $\pm$ 8 m~s$^{-1}$) should be detectable 
\citep{2007ApJ...655..550G},  allowing us to check the alignment
between the planetary orbit and the stellar spin axis.

\section*{Acknowledgments}
The WASP consortium comprises the Universities of Keele, Leicester, St. Andrews,
the Queen's University Belfast, The Open University and the Isaac Newton Group.
WASP-South is hosted by the South African Astronomical Observatory (SAAO) and we
are grateful for their support and assistance. Funding for WASP comes from
consortium universities and the UK's Science and Technology Facilities Council.
\bibliographystyle{mn2e}
\bibliography{DRA}

\begin{thebibliography}{}

\bibitem[\protect\citeauthoryear{{Bakos}, {L{\'a}z{\'a}r}, {Papp}, {S{\'a}ri}
  \& {Green}}{{Bakos} et~al.}{2002}]{2002PASP..114..974B}
{Bakos} G.~{\'A}.,  {L{\'a}z{\'a}r} J.,  {Papp} I.,  {S{\'a}ri} P.,    {Green}
  E.~M.,  2002, PASP, 114, 974

\bibitem[\protect\citeauthoryear{{Ben-Jaffel}}{{Ben-Jaffel}}{2007}]{2007ApJ...%
671L..61B}
{Ben-Jaffel} L.,  2007, ApJ, 671, L61

\bibitem[\protect\citeauthoryear{{Blackwell} \& {Shallis}}{{Blackwell} \&
  {Shallis}}{1977}]{1977MNRAS.180..177B}
{Blackwell} D.~E.,  {Shallis} M.~J.,  1977, MNRAS, 180, 177

\bibitem[\protect\citeauthoryear{{Bouvier}}{{Bouvier}}{1997}]{1997MmSAI..68..8%
81B}
{Bouvier} J.,  1997, Mem. Soc. Astron. Ital., 68, 881

\bibitem[\protect\citeauthoryear{{Burrows}, {Hubeny}, {Budaj} \& {Hubbard}}{{Burrows}
  et~al.}{2007}]{2007ApJ...661..502B}
{Burrows} A.,  {Hubeny} I.,    {Budaj} J., {Hubbard} W.~B., 1997, A\&A, 318, 841

\bibitem[\protect\citeauthoryear{{Castelli}, {Gratton} \& {Kurucz}}{{Castelli}
  et~al.}{1997}]{1997A&A...318..841C}
{Castelli} F.,  {Gratton} R.~G.,    {Kurucz} R.~L.,  1997, A\&A, 318, 841

\bibitem[\protect\citeauthoryear{{Collier Cameron} \& {et al.}}{{Collier
  Cameron} {et al.}}{2006}]{2006MNRAS.373..799C}
{Collier Cameron} A. et al., 2006, MNRAS, 373, 799

\bibitem[\protect\citeauthoryear{{Collier Cameron} \& {et al.}}{{Collier
  Cameron} {et al.}}{2007}]{2007MNRAS.380.1230C}
{Collier Cameron} A. et al., 2007, MNRAS, 380, 1230

\bibitem[\protect\citeauthoryear{{Fortney}, {Marley} \& {Barnes}}{{Fortney}
  et~al.}{2007}]{2007ApJ...659.1661F}
{Fortney} J.~J.,  {Marley} M.~S.,    {Barnes} J.~W.,  2007, ApJ, 659, 1661

\bibitem[\protect\citeauthoryear{{Fortney}, {Saumon}, {Marley}, {Lodders} \&
  {Freedman}}{{Fortney} et~al.}{2006}]{2006ApJ...642..495F}
{Fortney} J.~J.,  {Saumon} D.,  {Marley} M.~S.,  {Lodders} K.,    {Freedman}
  R.~S.,  2006, ApJ, 642, 495

\bibitem[\protect\citeauthoryear{{Gaudi} \& {Winn}}{{Gaudi} \&
  {Winn}}{2007}]{2007ApJ...655..550G}
{Gaudi} B.~S.,  {Winn} J.~N.,  2007, ApJ, 655, 550

\bibitem[\protect\citeauthoryear{{Girardi}, {Bressan}, {Bertelli} \&
  {Chiosi}}{{Girardi} et~al.}{2000}]{2000A&AS..141..371G}
{Girardi} L.,  {Bressan} A.,  {Bertelli} G.,    {Chiosi} C.,  2000, A\&AS, 141,
  371

\bibitem[\protect\citeauthoryear{{Gray}}{{Gray}}{1992}]{1992oasp.book.....G}
{Gray} D.~F.,  1992, {The Observation and Analysis of Stellar Photospheres}.
Cambridge University Press

\bibitem[\protect\citeauthoryear{{Guillot}, {Santos}, {Pont}, {Iro}, {Melo} \&
  {Ribas}}{{Guillot} et~al.}{2006}]{2006A&A...453L..21G}
{Guillot} T.,  {Santos} N.~C.,  {Pont} F.,  {Iro} N.,  {Melo} C.,    {Ribas}
  I.,  2006, A\&A, 453, L21

\bibitem[\protect\citeauthoryear{{Harrington}, {Luszcz}, {Seager}, {Deming} \&
  {Richardson}}{{Harrington} et~al.}{2007}]{2007Natur.447..691H}
{Harrington} J.,  {Luszcz} S.,  {Seager} S.,  {Deming} D.,    {Richardson}
  L.~J.,  2007, Nature, 447, 691

\bibitem[\protect\citeauthoryear{{Mandushev} \& {et al.}}{{Mandushev} {et
  al.}}{2005}]{2005ApJ...621.1061M}
{Mandushev} G. et al., 2005, ApJ, 621, 1061

\bibitem[\protect\citeauthoryear{{Mandushev} \& {et al.}}{{Mandushev} {et
  al.}}{2007}]{2007ApJ...667L.195M}
{Mandushev} G. et al., 2007, ApJ, 667, L195

\bibitem[\protect\citeauthoryear{{McCullough}, {Stys}, {Valenti}, {Fleming},
  {Janes} \& {Heasley}}{{McCullough} et~al.}{2005}]{2005PASP..117..783M}
{McCullough} P.~R.,  {Stys} J.~E.,  {Valenti} J.~A.,  {Fleming} S.~W.,  {Janes}
  K.~A.,    {Heasley} J.~N.,  2005, PASP, 117, 783

\bibitem[\protect\citeauthoryear{{O'Donovan} \& {et al.}}{{O'Donovan} {et
  al.}}{2006}]{2006ApJ...651L..61O}
{O'Donovan} F.~T. et al., 2006, ApJ, 651, L61

\bibitem[\protect\citeauthoryear{{Pollacco} \& {et al.}}{{Pollacco} {et
  al.}}{2006}]{2006PASP..118.1407P}
{Pollacco} D.~L. et al., 2006, PASP, 118, 1407

\bibitem[\protect\citeauthoryear{{Pollacco} \& {et al.}}{{Pollacco} {et
  al.}}{2007}]{WASP-3b}
{Pollacco} D.~L. et al., 2007, MNRAS, accepted, arXiv:0711.0126

\bibitem[\protect\citeauthoryear{{Pont} \& et al.}{{Pont}
  et~al.}{2007}]{2007arXiv0710.5278P}
{Pont} F.  et al., 2007, A\&A, submitted, arXiv:0710.5278

\bibitem[\protect\citeauthoryear{{Queloz} \& {et al.}}{{Queloz} {et
  al.}}{2001}]{2001A&A...379..279Q}
{Queloz} D. et al., 2001, A\&A, 379, 279

\bibitem[\protect\citeauthoryear{{Sestito} \& {Randich}}{{Sestito} \&
  {Randich}}{2005}]{2005A&A...442..615S}
{Sestito} P.,  {Randich} S.,  2005, A\&A, 442, 615

\bibitem[\protect\citeauthoryear{{Udalski}, {Szymanski}, {Kaluzny}, {Kubiak} \&
  {Mateo}}{{Udalski} et~al.}{1992}]{1992AcA....42..253U}
{Udalski} A.,  {Szymanski} M.,  {Kaluzny} J.,  {Kubiak} M.,    {Mateo} M.,
  1992, Acta Astron., 42, 253

\bibitem[\protect\citeauthoryear{{Vidal-Madjar}, {Lecavelier des Etangs},
  {D{\'e}sert}, {Ballester}, {Ferlet}, {H{\'e}brard} \& {Mayor}}{{Vidal-Madjar}
  et~al.}{2003}]{2003Natur.422..143V}
{Vidal-Madjar} A.,  {Lecavelier des Etangs} A.,  {D{\'e}sert} J.-M.,
  {Ballester} G.~E.,  {Ferlet} R.,  {H{\'e}brard} G.,    {Mayor} M.,  2003,
  Nature, 422, 143

\bibitem[\protect\citeauthoryear{{Vidal-Madjar}, {Lecavelier des Etangs},
  {D{\'e}sert}, {Ballester}, {Ferlet}, {H{\'e}brard} \& {Mayor}}{{Vidal-Madjar}
  et~al.}{2008}]{2008arXiv0802.0587V}
{Vidal-Madjar} A.,  {Lecavelier des Etangs} A.,  {D{\'e}sert} J.-M.,
  {Ballester} G.~E.,  {Ferlet} R.,  {H{\'e}brard} G.,    {Mayor} M.,  2008,
  ApJL, submitted, arXiv:0802.0587

\bibitem[\protect\citeauthoryear{{Weldrake}, {Bayliss}, {Sackett}, {Tingley},
  {Gillon} \& {Setiawan}}{{Weldrake} et~al.}{2007}]{2007arXiv0711.1746W}
{Weldrake} D.~T.~F.,  {Bayliss} D.~D.~R.,  {Sackett} P.~D.,  {Tingley} B.~W.,
  {Gillon} M.,    {Setiawan} J.,  2007, ApJL, accepted, arXiv:0711.1746

\bibitem[\protect\citeauthoryear{{Wilson} \& {et al.}}{{Wilson} {et
  al.}}{2007}]{WASP-4b}
{Wilson} D.~M. et al., 2007, ApJL, accepted, arXiv:0801.1509

\end{thebibliography}

\label{lastpage}
\end{document}